\documentclass[referee]{aa} 
\usepackage{graphicx} 
\usepackage{txfonts}
\begin{document}
   \title{Blue Straggler Stars in Galactic Open Clusters and
          the effect of field star contamination}

   \author{Giovanni Carraro
          \inst{1}
          \and
          Ruben A. V\'azquez
          \inst{2}
          \and
          Andr\'e Moitinho
          \inst{3}
          }

   \offprints{Giovanni Carraro}

   \institute{ESO, Alonso de Cordova 3107, Vitacura, Santiago, Chile
\thanks{on leave from Dipartimento di Astronomia, Universit\`a di Padova,
                 Vicolo Osservatorio 2, I-35122 Padova, Italy}\\
              \email{gcarraro@eso.org}
         \and
               Facultad de Ciencias Astron\'omicas y Geof\'{\i}sicas de la
                 UNLP, IALP-CONICET, Paseo del Bosque s/n, La Plata,
                 Argentina\\
             \email{rvazquez@fcaglp.unlp.edu.ar}
          \and
           SIM/IDL, Faculdade de Ci\^encias da Universidade de
Lisboa, Ed. C8, Campo Grande, 1749-016 Lisboa, Portugal\\
\email{andre@sim.ul.pt}  }

   \date{Received ...; accepted ....}

 
  \abstract
   {We investigate the distribution of Blue Straggler
  stars in the field of three open  star clusters.}
   {The main purpose is to highlight the crucial role played by
  general Galactic disk fore-/back-ground field stars, which
  are often located in the same region of the Color Magnitude
  Diagram as Blue Straggler stars.}
   {We analyze photometry taken from the literature of 3 open clusters of
  intermediate/old age rich in Blue Straggler stars, and which are 
  projected in the direction of the
  Perseus arm, and study their spatial distribution and the Color
  Magnitude Diagram.}
   {As expected, we find that a large portion of the Blue Straggler
  population in these clusters are simply young field stars belonging
  to the spiral arm. This result has important consequences on the theories
  of the formation and statistics of Blue Straggler stars in different
  population environments: open clusters, globular clusters or dwarf galaxies.}
  {As previously emphasized by many authors, a detailed membership
  analysis is mandatory before comparing the Blue Straggler population
  in star clusters against theoretical models. Moreover, these
  sequences of young field stars (blue plumes) are potentially
  powerful tracers of Galactic structure which require further consideration.}


   \maketitle
%

\section{Introduction}
In the last few years blue straggler stars (BSS) have received 
considerable attention. 
These bright stars  appear above the turn off point
(TO) in the color magnitude diagram (CMD) of stellar populations, 
along the extension of the main sequence (MS), and are almost ubiquitous. 
In fact BSS have been detected in globular 
clusters (Sandage 1953, 
Ferraro et al. 2003), Dwarf Galaxies in the Local Group
(Momany et al. 2007), and Galactic open clusters (OCs) (Ahumada \& Lapasset
2007, De Marchi et al. 2006, Xin et al. 2007).
The formation mechanism of BSS in different environments is still
lively debated, and we refer the reader to the quoted papers for all the
details. \\
According to the recent analysis
by  Momany et al. (2007) Galactic open clusters,
although being less populous than globular clusters and dwarf galaxies,
seem to contain the largest percentage of BSS (see their Fig.~2).
This appears to imply that the formation mechanism of BSS depends on the
environment, being different in open and globular clusters and in
dwarf spheroidal galaxies. 
However, open clusters are embedded in the Galactic disk, and are
more affected by field star contamination than globulars.
The amount of contamination depends on the cluster position,
and can only be quantified by a detailed membership analysis, which
unfortunately is available for only a handful of clusters (see
Ahumada \& Lapasset 2007 for more details).\\
\noindent
A related problem in fact is that blue sequences in the CMD of dwarf
galaxies (the so called {\it blue plumes}) bear much similarity with
BSS sequences in open and globular clusters, but at the same time can
be made of young stars, residuals of recent star formation
episodes (Momany et al. 2007).\\
A remarkable case is that of Canis Major (CMa, Bellazzini et
al. 2004), a putative dwarf galaxy immersed in the third quadrant of
the Galactic disk.  The prominent {\it blue plume} visible in CMDs has
been considered to be composed of 1-2 Gyr old
stars and/or  BSS (Bellazzini et al. 2006) belonging to CMa.\\
However, in a series of papers (Carraro et al. 2005, Moitinho et
al. 2006, Pandey et al. 2006) similar sequences have been observed in
CMDs across the entire third Galactic quadrant, and have been argued -
based on a larger wavelength coverage - to be composed of young stars
belonging to previously poorly known spiral features. These different
views highlight the delicacy of interpreting CMDs and how the choice
of photometric bands may be critical.

\noindent
In this {\it Note} we investigate the BSS population in three OCs (NGC
7789, Berkeley 66 and Berkeley 70), located in the second Galactic
quadrant, where the spiral structure is well known. 
We draw attention on the effect of field star
contamination in cluster CMDs and to the need of detailed membership
studies as the observational foundation for robust theoretical
investigations of BSS formation mechanisms.  Moreover, we show that the
blue sequences like those found in the third Galactic Quadrant are
also present in the second quadrant, and can be in most cases ascribed
to spiral features.

   \begin{figure}
   \centering
   \includegraphics[width=\columnwidth]{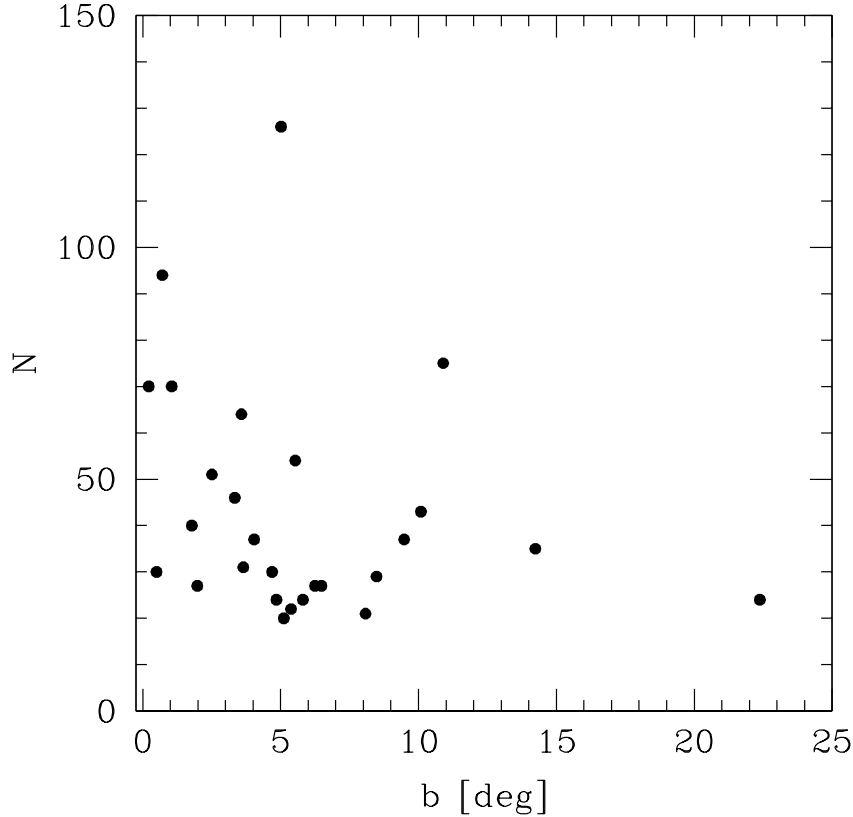}
   \caption{ Absolute number of BSS in the most BSS rich
open clusters (AL07) {\it vs} their Galactic latitude.}
    \end{figure}

\section{The role of field star contamination}
Our starting point is the new release of the BSS catalog by Ahumada \&
Lapasset (2007, hereinafter AL07).  We opted to use this compilation
amongst the various at disposal because it contains the most
up-to-date list of BSS candidates in OCs.  As the authors state,
accurate membership studies, based either on radial velocity or proper
motion, are unfortunately only available for a limited number of
clusters, and therefore it is only possible to provide BSS candidates
according to some less accurate criteria.  The fact that BSS
stars may occupy the same locus as field stars in the CMD is the main
difficulty one has to deal with when providing robust BSS statistics.

    To illustrate how crucial this contamination is, we consider the
    star clusters with the largest absolute populations of BSS listed
    in Table~6 of AL07.  From the sample, we remove Ruprecht~46, which
    - interestingly for the purpose of this {\it Note} - has been
    demonstrated to be a chance alignment of field stars (Carraro \&
    Patat 1995).  Yet, this random enhancement of field stars leads
    the entries of {\it clusters} having the
    largest relative population of BSS (see Table~7 of AL07).\\
    In Fig.~1, we plot the 28 entries of Table~6 of AL07 as a function
    of Galactic latitude. Clearly, there is a trend of having
    more BSS at lower Galactic latitudes, where field star
    contamination becomes more important.

\noindent
Recently, Momany et al. (2007) have shown that field star
contamination is the source of an artificially higher BSS fraction in
the globular clusters NGC~6717 (l = -11$^{o}$) and NGC~6838 (l =
-5$^{o}$) when compared to clusters of similar absolute magnitude 
but located at higher Galactic latitudes.  The same seems to occur for
Galactic open clusters.  In the AL07 catalog - but also in all the
other compilations ( Xin et al. 2007, De Marchi et al. 2007) - M~67 and
Berkeley~18 are two coeval and similar metallicity clusters, still
Berkeley~18 at (l,b) = (163.63,+5.02) has 4 times more BSS than M~67
at (l,b)=(215.69,+31.89).

\begin{table}
\caption{Fundamental parameters of the clusters under study, taken
  from WEBDA}
\fontsize{8} {10pt}\selectfont
\begin{tabular}{ccccccc}
\hline
\multicolumn{1}{c} {$Name$} &
\multicolumn{1}{c} {$l$} &
\multicolumn{1}{c} {$b$} &
\multicolumn{1}{c} {$Radius$} & 
\multicolumn{1}{c} {$(m-M)_V$} &
\multicolumn{1}{c} {$E(V-I)$} &
\multicolumn{1}{c} {$age$} \\
\hline
 & deg & deg  & arcmin &  &  & Gyr \\
\hline
NGC~7789    & 115.53 &  -5.38 & 14.0 & 12.5 & 0.27 & 1.7\\
Berkeley~66 & 139.43 &  +0.22 &  2.0 & 17.5 & 1.55 & 5.0\\
Berkeley~70 & 166.89 &  +3.58 &  2.5 & 14.6 & 0.60 & 4.7\\
\hline
\end{tabular}
\end{table}

\section{The spiral structure of the Milky Way}
The majority of OB type stars are located in spiral arms and form
sequences in the CMD blurred by differential reddening and distance
spread due to the patchy and irregular shape of the
arms. Still, these sequences are prominent.\\
In the last years we have provided evidence of such sequences
in the fields of open clusters in the Third Galactic Quadrant
(TGQ, Carraro et al. 2005; Moitinho et al. 2006), where no previously
accepted indications of spiral arms were present.  Similar results
have been obtained by Pandey et al. (2006) for a sample of clusters in
the 3GQ and Second Galactic Quadrant (SGQ).  

We searched for the presence of spiral arm traces 
lines-of-sight of clusters listed in Table~6 of AL07 to understand
whether BSS candidates could be mostly interpreted as young field
stars.  As an illustration, we consider here 3 cases: NGC~7789,
Berkeley~66 and Berkeley~70, located at (l,b) = (115.53,
-5.38),(139.43,+0.22) and (166.89,+3.58), respectively.  These are
three OCs with high relative populations of BSS (AL07, Table~7) and well
separated in the SGQ.  At odds with the case of the
TGQ, these three clusters are located in the SGQ along directions
where the presence of Perseus and of the Local arm are well
established (Russeil et al. 2007).  Moreover, according to the maps of
Burton (1985), one can detect HI emission along these directions due
to the prominent Local arm, which surrounds the Sun, and to the more
distant, detached, Perseus arm.

   \begin{figure}
   \centering
   \includegraphics[width=\columnwidth]{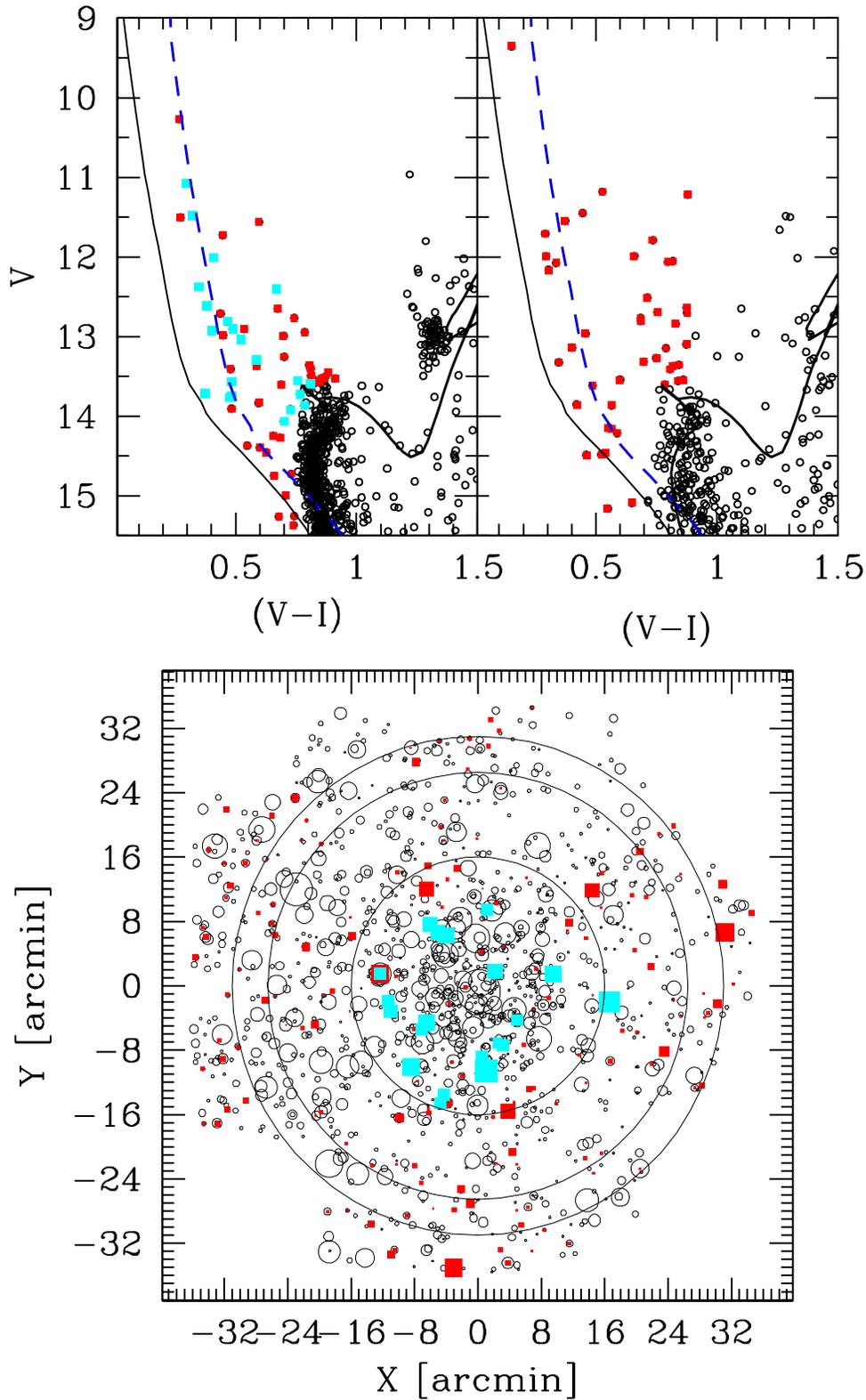}
   \caption{Location in the CMD of NGC~7789 of BBS and field stars.
     In the upper left panel we show CMDs of NGC7789 from Gim et
     al. (1998) for all stars brighter than V $\sim$ 15.50 and within the cluster radius
     (see Table~1). In the upper right panel, the CMD of an equal area field is shown
     for comparison. The cluster and field area are indicated in the lower panel, where
     the field of NGC~7789 covered by Gim et al. (1998) photometry is shown. In this
     panel the innermost circle indicates the cluster area, while the two outermost circles
     illustrate the field selection. This annulus have been chosen to have the same area and
     at the same time to be as far as possible distant from the cluster, to avoid contamination
     from NGC~7789 external halo stars.
     In all the panels filled squares (in red when color printed) refer to BSS stars, according
     to AL07 criteria (see text). Besides, different color filled squares (in cyan when color printed) are used 
     for the BSS members of NGC~7789.
     In the CMDs, the two solid lines are an isochrone and a ZAMS drawn for the value
     of the fundamental parameters of NGC~7789 (see table~1). On the other hand, the dashed
     line (blue when printed in color) has been drawn for the typical distance of the Perseus
     arm in the NGC~7789 direction.}
    \end{figure}

\subsection{NGC 7789}
At l=115$^{o}$ (the Galactic longitude of NGC~7789), the maps of Burton (1985)
clearly indicate that the Perseus arm is detached from the Local arm,
and show HI emission down to at least b = -6$^{o}$.  This implies that
contamination from young stars in the Perseus arm
can be important in the field of NGC~7789.\\
In the upper panels of Fig.~2 we show the CMD of NGC~7789 (left panel) and of 
the surrounding field (right panel).
In the lower
panel we plot the stellar spatial distribution, and illustrate with a circle
14 arcmin wide (the innermost one) the cluster region. An annulus far from the cluster
and having the same area as the clusters is indicated with the two outermost circles.
In particular,
together with all the observed stars, we
indicate with solid symbols (red squares when printed in color)
the spatial distribution of the candidate
BSS, as selected in each CMD. As for the selection, we adopted the
same criteria as AL07 (see their Fig.~1): The solid isochrone and ZAMS
(thick solid line) are drawn using the fundamental parameters of NGC~7789 (see Table~1) to
show the region of the CMD where BSS candidates are to be searched
for.  We are aware that for this specific cluster a
number of BSS (22) have been found to be cluster members, and we
indicate them  with magnitude-sized squares (cyan when printed in color)
in Fig.~2.
However, for the aim of this {\it Note}, what is actually
relevant is the number of candidate BSS ({\it lacking a membership
analysis in most cases, all stars in the appropriate region of
the CMD are routinely considered to be BSS candidates}), and therefore
we did not remove the members BSS from the candidate
sample. \\
The first important fact to highlight is that candidate BSS are evenly
distributed across the field.  In the CMDs in NGC~7789 area (Fig~2, upper
panels) we recognize a vertical sequence right above the TO, and an
almost parallel, detached, bluer sequence. The BBS (see also
Fig.~4 in AL07) do not follow closely none
of these sequences, but are significantly spread in color.\\
\noindent
While the sequence right above the TO is most probably due to a mix of
binary stars and nearby field stars, the bluer sequence 
resembles the {\it blue plumes} found in dwarf galaxies
(Momany et al 2007), or in the background of open clusters
in the  3GQ (Carraro et al. 2005).\\
The bulk of these blue stars does not follow the solid ZAMS
(compatible with stars at the same distance and reddening of NGC
7789), but lies red-ward.  We suggest that this is a sequence of more
distant and reddened young stars, similar to a {\it blue plume}.  
The
dashed ZAMS (blue when printed in color) 
we over-imposed on it would in fact imply a distance of
about 3 kpc, significantly larger than the distance of NGC~7789 (less
than 2 kpc).\\
Along the {\it blue plume} we count about 34  stars.
From AL07 we know that NGC 7789 harbors 22 BSS
which are almost entirely concentrated 
within the cluster radius (see the bottom panel map). 
However, only a fraction of them (13) lie close to this blue sequence.\\
\noindent
We conclude that a substantial fraction of the stars 
in the {\it blue plume} are young field stars 
located in the southern tail of the Perseus arm along the cluster line of sight,
as evident also from the BSS candidates position in the field CMD (upper right panel).
This result is supported by
the HI radial velocity in the range of -65$:$-75 km/sec (Burton 1985), somewhat
larger than NGC~7789  ( -54.9$\pm$0.12 km/sec, Gim et al. 98).\\
Moreover, the NGC~7789 isochrone distance is in fact 
smaller than 2.0 kpc, 
while the distance to the HII regions associated to
the arm (S163, S164, S166, and S170 for
instance)  at the same longitude of the cluster
are larger (2.5$--$2.9 kpc, Russeil et al 2007), and  compatible
with the {\it blue plume}.

\subsection{Berkeley~66}
This is a distant open cluster, located well beyond the Perseus arm,
but very close to the plane, and for that reason very reddened (Phelps
\& Janes 1996, Guarnieri \& Carraro 1997, Villanova et al. 2005).  
It represents an interesting
case since it lies along the line of sight where recently Xu et
al. (2006) have estimated the distance to the Perseus arm to be about
2 kpc.  In Fig.~3 we show the CMDs of Berkeley~66.  The photometry is
taken from Phelps \& Janes (1996), who covered a $5^{\prime}.1 \times
5^{\prime}.1$ field around the cluster's nominal center and found that
it has a radius smaller than 2 arcmin.  This field of view is small
and therefore we do not expect to find many stars associated to the
Perseus arm, which lies much closer than the cluster.\\
\noindent
In the left panel we consider only stars within the nominal cluster
radius (2 arcmin, see Table~1), as defined by the circle in the lower panel map. 
We super-imposed an isochrone for the values of the basic
parameters as in Table~1.  A ZAMS for this reddening and distance modulus
is also plotted with the same symbols to guide the eye and indicate
the region where BSS (filled symbols) are to be searched for,
according to the AL07 criteria. The same representation is used
in the field CMD (upper right panel), which contains all the stars outside
the cluster area. This field -limited by the outermost circle in the map- 
has the same area of the cluster region.
The spatial distribution of BSS candidates in the cluster and field
is show in the bottom panel, where the area covered
by Phelps \& Janes photometry is shown. 

Within the cluster radius (2$^{\prime}$) we count 63 BSS candidates (AL07 indicate
70, but possibly used a different radius).  All stars 2.5 mag brighter
than the TO are taken as BSS candidates.  However, looking at the
field CMD, the color and magnitude distribution of BSS candidates and
the field stars are similar. In fact, we count in the field 68
BSS candidates. It is therefore likely that a significant number ofBSS
candidates are just field stars.

Although located behind the Perseus arm, no obvious traces of a tight young
star sequence associated to the arm ({\it blue plume}) are found in the field
of Berkeley~66.
This is most probably due to the small field of view of the
observations together with the highly irregular absorption
and/or to the patchy structure of the arm.

   \begin{figure}
   \centering
   \includegraphics[width=\columnwidth]{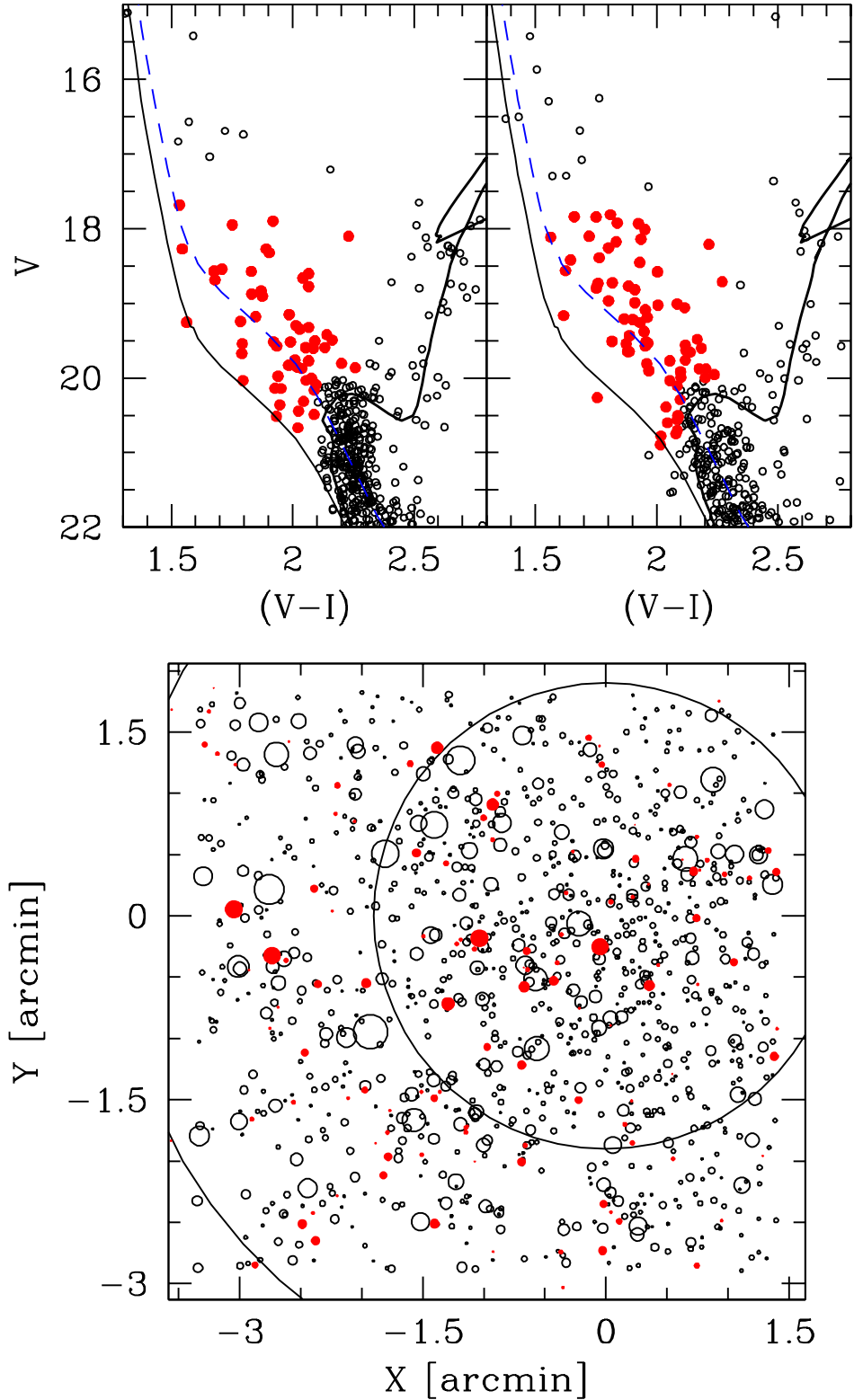} 
   \caption{Upper panels: CMD of Berkeley~66 for cluster (left)  and
   field (right)
   regions, as defined in the lower panel, where the innermost circle is drawn 
   adopting 2.0 arcmin for the cluster radius (see Table~1). The region outside the cluster
   area and limited by the outermost circle has the same area of the cluster region.
   Solid symbols
   (red when printed in color) indicate BSS candidates. The solid lines are
   a ZAMS and an isochrone for the most accepted values of the cluster
   fundamental parameters (as in Table~1), whereas the dashed line (blue when 
   printed in color) indicates the expected location of young stars at the 
   distance of the Perseus arm.}
    \end{figure}

   \begin{figure}
   \centering
   \includegraphics[width=\columnwidth]{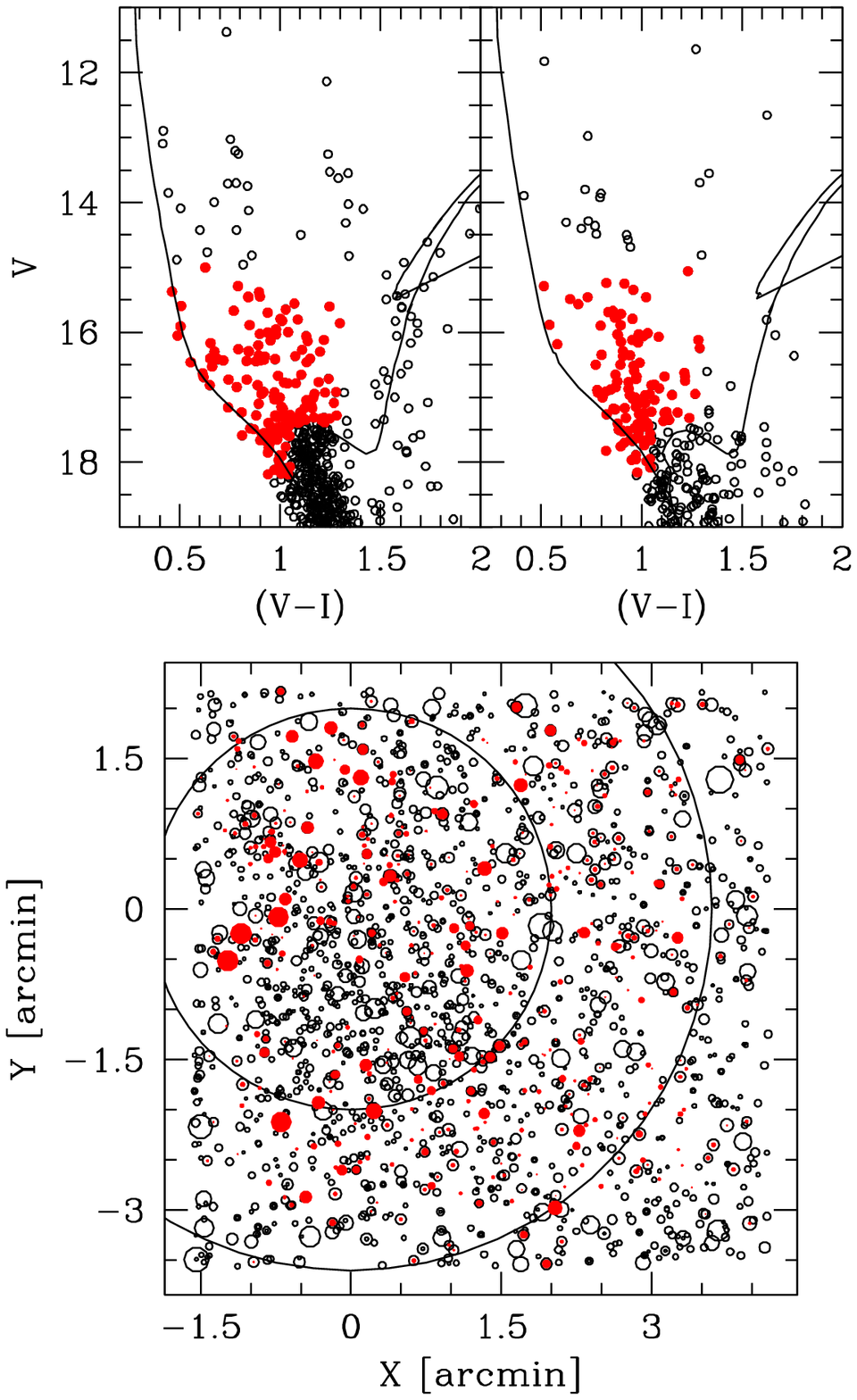}
   \caption{Upper panels: CMD of Berkeley~70 for cluster (left)  and
   field (right)
   regions, as defined in the lower panel, where the innermost circle is drawn 
   adopting 2.5 arcmin for the cluster radius (see Table~1). The region outside the cluster
   area and limited by the outermost circle has the same area of the cluster region.
   Solid symbols
   (red when printed in color) indicate BSS candidates. The solid lines are
   a ZAMS and an isochrone for the most accepted values of the cluster
   fundamental parameters (as in Table~1).}
    \end{figure}

\subsection{Berkeley~70}
This is another old star cluster reported to be very rich in BSS. AL07
list 64 BSS candidates, virtually all the stars above the TO.
Although Berkeley~70 is relatively high above the Galactic plane
(b=+3.6$^{o}$, Z approximately 250 pc), Burton (1985) has detected
considerable HI emission toward its direction, which indicates that
the Perseus arm is significantly thick, reaching up to b = +8$^{o}$
degrees.  We take the photometry of Berkeley~70 from Ann et
al. (2002), who performed a study of a $5^{\prime}.8 \times
5^{\prime}.8$ field centered on the cluster.  The CMD is presented in
Fig.~4 for the cluster (upper left panel) aqnd field (upper right panel) region,
together with the spatial distribution of all the stars and the
candidate BSS (filled red circles when printed in color) 
in  the lower panel map.
Here the innermost circle corresponds to the cluster radius ($\sim$ 2.5 arcmin),
whereas the region outside the cluster area and limited by the outermost circle
identifies an equal area stellar field.\\
In the CMDs we
over-imposed an isochrone and a ZAMS (solid lines) using the values
from Table~1, and marked BSS candidates (following AL07) with filled
symbols (red when printed in color).
\noindent
The upper part of the CMD is characterized by a number of blue stars
close to the prolongation of the cluster's ZAMS, and another more
scattered sequence right above the TO.  The first of the two is a
sequence of young stars at about the same distance of the cluster
($\sim$ 4.0 kpc), similar to the distance to the Perseus arm in this
direction (Russeil et al. 2007), and therefore a fraction of these
blue stars are likely to belong to the arm. Therefore, for the specific
case of Berkeley~70, the Perseus arm young stars have roughly the same
distance as the cluster, and for this reason we do not plot any additional
ZAMS.\\
As an additional confirmation of their field stars nature,
these stars are
evenly distributed across the observed field and - according to the
cluster radius (2.5 arcmin) - only a fraction of them lies within
the cluster.  The other, redder and more scattered sequence, is
probably due to contamination of field stars near to the Sun,
including stars in the Local arm. 
Beside, we  note that other bright stars (brighter than the upper limit
for BSS candidates according to AL07) lie close to this ZAMS, emphasizing
their nature of young field stars associated with the arm.\\ 
Also in this case, the number and distribution of BSS candidates
in the cluster and field CMDs are very similar.

\noindent
We therefore conclude that the distribution of young blue stars in
Berkeley~70 is that expected from the location of the cluster,
right above and at the same distance of the Perseus arm.

\section{Conclusions}
The aim of this {\it Note} was twofold. On one side, we wanted
to stress once again how derived statistics of BSS stars in Galactic
OCs is crucially dependent on a proper accounting of field star
contamination. Specifically, this paper addresses the effect of
contamination introduced by the early type stellar component of spiral
arms. Trends like the anti-correlation between absolute magnitude
and BSS frequency (Piotto et al. 2004) must be reconsidered in the
light of accurate radial velocity or proper motion-based
memberships. Only this way the derived BSS population
in OCs would be  statistically meaningful. We are in the process of starting
such observational effort for a number of selected open clusters.\\
On the other side, we have shown how the bulk of the stars populating
the upper part of three representative open clusters (NGC 7789,
Berkeley 66 and Berkeley 70) in the SGQ is mostly dominated by young
stars located nearby the Sun - in the {\bf Local} arm - and in the Perseus
arm, which lies in front of (for Berkeley 66 and Berkeley 70), or
beyond (for NGC 7789) these clusters.  This was possible since we
already know the existence, distance and location of the Perseus arm.
This provides a confirmation that similar sequences found in the CMDs
of stellar fields in other regions of the Galaxy can be used as spiral
arm tracers (Carraro et al.  2005, Moitinho et al. 2007, V\'azquez et
al. 2007).

\begin{acknowledgements}
This study made use of SIMBAD and WEBDA. A.M. acknowledges support from
FCT (Portugal) through grant PDCT/CTE-AST/57128/2004. 
R.A.V. acknowledges the financial support from the CONICET PIP 5970. 
We thank the anonymous
referee for her/his suggestions which helped us to improve
the paper presentation.
\end{acknowledgements}

\end{document}